\documentstyle[12pt]{article}

\def\B{{\cal B}}
\def\H{{\cal H}}
\def\h{{\bf h}}
\def\L{{\cal L}}
\def\O{{\cal O}}
\def\P{{\cal P}}
\def\PP{\hat{\cal P}}
\def\p{\bf p}

\def\R{{\bf R}}
\def\x{\bf x}
\def\y{\bf y}
\let\@=\|
\def\?#1?{\left\@#1\right\@}
\def\!#1!{\left\@#1\right\@}
\def\<#1|#2>{\left<#1\vphantom{#2}\right|\left.\vphantom{#1}#2\right>}
\def\|#1|{\left|#1\right|}
\def\(#1){\left(#1\right)}
\def\[#1]{\left\lbrace #1\right\rbrace}

\begin{document}
\date{June 1994}
\title{Cluster Estimates and Analytic Wavefunctions}
\author{D.~R.~Davidson\\Department of Physics\\
University of California, Berkeley}
\maketitle
\begin{abstract}
The Tomita-Takesaki modular theory is used to establish a cluster estimate
extending and modifying that of Thomas and Wichmann [11], so as to extend it to
regions within which the relevant observables are not necessarily spacelike
separated.  This sort of estimate is then applied to the case of a massive free
field, to show that wavefunctions localized in a certain sense are analytic
functions of momentum.
\end{abstract}
\bigskip

\section{Introduction}
In [11] Thomas and Wichmann used the Tomita-Takesaki modular theory to
demonstrate one version of the well-known [1,6] exponential decay of matrix
elements of products of spacelike separated operators, or equivalently of
matrix elements of spatial translation operators between localized vectors.
These sorts of results are called cluster estimates because they express the
decay of correlations between clusters of observables as the spatial separation
between clusters increases.  Here we present an extended version of their
estimate, one which extends to the region within which the observables are not
strictly spacelike separated.  This version uses localized vectors that have
their high-energy behavior tamed by multiplication by an exponential cutoff
$e^{-\rho H}$, where $H$ is the Hamiltonian and $\rho$ is a constant
representing the size of their region of localization.  Once we have
established this sort of estimate, we will present an application:  a proof
that certain localized free-field wavefunctions are analytic functions of
momentum.

As in [11], the general framework is either that of a relativistic quantum
field theory [10], or else that of a relativistic quantum system of von Neumann
algebras of local observables [7].  In both cases a central feature of the
theory is a map from certain subsets $\O$ of Minkowski space to algebras of
operators on a Hilbert space $\H$:  in the first case, algebras $P(\O)$ of
unbounded averaged field operators defined on a common domain; in the second,
von Neumann algebras $\B(\O)$ of bounded operators considered to be observable
within the region $\O$.  This map is such that $P(\O_1)\subset P(\O_2)$, or
$\B(\O_1)\subset\B(\O_2)$, whenever $\O_1\subset\O_2$.

In both cases the Hilbert space $\H$ carries a strongly continuous unitary
representation $U_\lambda$ of the universal covering group $\PP$ of the
Poincar\'e group $\P$, such that if $\lambda$ is any Poincar\'e transformation,
and $\lambda\O$ is the image of $\O$ under $\lambda$, then
$U_\lambda P(\O) U^{-1}_\lambda=P(\lambda\O)$, or
$U_\lambda\B(\O) U^{-1}_\lambda=\B(\lambda\O)$.  The spectral condition
requires that the representation be such that the spectrum of the translations
is confined to the forward light cone.  For the purposes of these estimates, it
will be necessary to assume that there is a mass gap:  apart from the unique
Poincar\'e-invariant vacuum vector $\Omega$, the spectrum of the translations
is supported above the mass hyperboloid with mass $m_0$.

In both frameworks it is possible to define vectors localized in $\O$
in a certain sense, namely that they are produced by the application of
self-adjoint local (bounded or unbounded) operators to the vacuum;
in other words, $P(\O)^{\rm sa}\Omega$ or $\B(\O)^{\rm sa}\Omega$.  If the
quantum field theory and the system of local algebras are locally associated in
an appropriate sense [5] then these two sets will have the same closure, a
closed real-linear manifold
$R(\O)=\overline{P(\O)^{\rm sa}\Omega}=\overline{\B(\O)^{\rm sa}\Omega}$.
The Reeh-Schlieder principle implies that $R(\O)+iR(\O)$ is dense in $\H$; we
will consider the vectors in $R(\O)+iR(\O)$ to be localized in $\O$, and it is
with these localized vectors that we will primarily be concerned.  They are
not, of course, strictly localized [8]; if they were, we would not be
discussing the decay of their inner products at spacelike separations.
However, they are natural analogues of the strictly localized wavefunctions of
non-relativistic quantum mechanics, and they are natural objects of study
within these frameworks.

There are several facts available to us about these localized vectors.  First,
we have the information provided by the principle of locality, that observables
localized in spacelike separated regions commute.  Since we wish to allow for
the possibility of fermionic fields and local operators, which are not strictly
observable, we must generalize this to include anticommutation.  For this we
will use the device of Bisognano and Wichmann [2], and define $Y^z=ZYZ^{-1}$,
where $Z=(I+iU_0)/(1+i)$, and $U_0$ represents the rotation by $2\pi$ about any
axis.  Then the condition of locality simply states that $[X,Y^z]=0$ whenever
$X$ and $Y$ are localized in spacelike separated regions.  If $X$ and $Y$ are
self-adjoint, this implies that the inner product
$\<X\Omega|Y^z\Omega>=\<X\Omega|ZY\Omega>$ is purely real.  If $\O_1$ and
$\O_2$ are spacelike separated, it follows that $\<\psi|\phi>$ must be real
whenever $\psi\in R(\O_1)$ and $\phi\in ZR(\O_2)=R(\O_2)$.  The principle of
duality goes further, and requires that $R(\O^c)$ consist precisely of those
vectors whose inner products with all vectors in $R(\O)$ are real, where $\O^c$
is the spacelike complement of the region $\O$.  This can be (and usually is)
regarded as a maximality requirement on the local algebras, and we will
hereafter assume it.  (Rieffel [9] shows that the version stated here implies
the usual one.)

The principle of special duality [2] goes further still, and requires that the
real-linear manifolds $R(W_R)$ and $R(W_L)$ for the wedge regions
$W_R=\[x_3>\|t|]$ and $W_L=\[x_3<-\|t|]$ be specifically described in a certain
way.  It is known from the Tomita-Takesaki modular theory that they can always
be given as $R(W_R)=\[\psi\Bigm|\psi=J\Delta^{1/2}\psi]$ and
$R(W_L)=\[\psi\Bigm|\psi=J\Delta^{-1/2}\psi]$ for some modular operators $J$,
$\Delta^{1/2}$, where $\Delta$ is a positive (unbounded) operator and $J$ is an
antilinear involution.  Special duality specifies the form of these modular
operators as follows.  Let $V_3(t)=V(t,\hat{\x}_3)$ be the representatives of
the velocity transformations in the $\hat{\x}_3$ direction, whose natural
action on Minkowski space is given by the matrix
\begin{equation}
M(t)=\pmatrix{\cosh t &0&0&\sinh t\cr 0&1&0&0\cr 0&0&1&0\cr
\sinh t&0&0&\cosh t\cr}.
\end{equation}
To produce the modular operator $\Delta^{1/2}$ for the right wedge, it is
necessary to perform an analytic continuation:  let the complex variable $t$
now be $\sigma+i\tau$, and let $D(i\tau)$ be the domain of $V_3(t)$, which
depends only on $\tau$, and is such that $D(i\tau')\supset D(i\tau)$ whenever
$0\leq \tau'\leq\tau$ or $0\geq\tau'\geq\tau$.  Then the modular operator for
the right wedge is $\Delta^{1/2}=V_3(i\pi)$, with domain
$D(i\pi)=R(W_R)+iR(W_R)$, and the modular operator for the left wedge is
$\Delta^{-1/2}=V_3(-i\pi)$, with domain $D(-i\pi)=R(W_L)+iR(W_L)$.  The modular
conjugation $J$ is given by $J=ZU\Theta$, where $Z$ is as above, $U$ represents
a rotation by angle $\pi$ about the 3-axis, and $\Theta$ is the TCP operator.
The result of Bisognano and Wichmann [2] is that this relation holds in the
case of algebras produced from Wightman fields, and thus also in the case of
local algebras locally associated to Wightman fields.  We will assume hereafter
that special duality does indeed hold.

Notice that $R(W_R)$ and $R(W_L)$ depend on the TCP operator $\Theta$, but if
we are interested only in $R(W_R)+iR(W_R)$ and $R(W_L)+iR(W_L)$, these depend
only on the representation of the Lorentz group.  If two systems---for example,
an interacting field theory and its asymptotic free theory---share a common
representation of $\PP$, then they have the same localized vectors for any
wedge region.  This does not necessarily imply anything about the localized
vectors for other regions, such as the double-cone regions; we know that
$R(K)\subset R(W)$ if $K\subset W$, but there is no reason to think that we can
compute $R(K)$ even given all of the $R(W)$.  In particular, it is not the case
that $R(K)$ is equal to the intersection of the $R(W)$ for all $W\supset K$.
This will become clearer in the third section, when we discuss $R(K)$ in the
particular case of the free fields.

\section{A Cluster Estimate}

The form of the cluster estimate due to Fredenhagen [6] was originally stated
in terms of bounded operators $A$ and $B$, which we can take to be localized in
regions separated by a spacelike distance $s$; then his estimate was
\begin{equation}
\|{\<\Omega|AB\Omega>-\<\Omega|A\Omega>\<\Omega|B\Omega>}|
\leq e^{-sm_0}\sqrt{\!A^*\Omega!\!A\Omega!\!B^*\Omega!\!B\Omega!}.
\end{equation}
This form explicitly exhibits the cluster decomposition nature of the estimate,
but it will be convenient hereafter to use a slightly different form.  Let $F$
be the projection onto $\[\Omega]^\perp$, the orthogonal complement of the
vacuum; $F$ will commute with $\PP$, and $F\psi$ will be a localized vector if
and only if $\psi$ is.  Then we may write the left-hand side of (2) as
$\|{\<A^*\Omega|FB\Omega>}|$, and the estimate will appear as a bound on the
matrix elements of inner products of the localized vectors $FA^*\Omega$ and
$FB\Omega$.  Thomas and Wichmann [11] established the following estimate:  let
$\psi\in D(-i\pi/2)$ and $\phi\in D(i\pi/2)$; then for $s\geq 0$,
\begin{equation}
\|{\<\psi|T(s\hat{\x}_3)F\phi>}|\leq
e^{-sm_0}\!FV_3(-i\pi/2)\psi!\!FV_3(i\pi/2)\phi!,
\end{equation}
where $T(x)$ is the representative of the translation by $x$.  This can be
placed in the form of (2) by noting that if $\phi=B\Omega$, where $B$ is local
to the right wedge, then $\phi\in D(i\pi)\supset D(i\pi/2)$, and
\begin{equation}
\!FV(i\pi/2)\phi!^2=\<FB\Omega|FV(i\pi)B\Omega>
=\<FB\Omega|FJB^*\Omega>\leq\!B\Omega!\!B^*\Omega!;
\end{equation}
likewise if $\psi=A^*\Omega$, where $A$ is local to the left wedge, then
$\psi\in D(-i\pi)\supset D(-i\pi/2)$, and
\begin{equation}
\!FV(-i\pi/2)\psi!^2=\<FA^*\Omega|FV(-i\pi)A^*\Omega>
=\<FA^*\Omega|FJA\Omega>\leq\!A\Omega!\!A^*\Omega!.
\end{equation}
These estimates are useful in case the two regions of localization are
separated by some positive distance $s$, but in some cases one wishes to have
an estimate for matrix elements like those of (3) that will also cover the case
in which the regions of localization overlap.  That is what we will provide
here.

We find that if we multiply $\psi$ and $\phi$ by an exponential cutoff in
energy $e^{-\rho H}$, a modification which presumably improves these vectors'
high-energy behavior, we can in fact produce estimates that do not depend on
strict spacelike separation.  Cutoffs of this sort have been considered in
connection with nuclearity requirements [3], in an attempt to characterize
models with reasonable particle interpretations, but it is not immediately
clear why they should appear in cluster estimates.  Nevertheless, we will see
that just such a cutoff provides the essential element in eliminating the
requirement of spacelike separation.

{}From the definition of the Poincar\'e group we find that
$V_3(t)T(x)V_3(t)^{-1}=T(M(t)x)$ for real $t$ and $x$, and this relation can be
extended by analytic continuation on the appropriate domains, bearing in mind
that $V_3(t)^{-1}=V_3(-t)=V_3(t^*)^\dagger$, and that
$T(i\rho\hat{\x}_0)=e^{-\rho H}$ is bounded for $\rho\geq 0$.  We can then
establish the following:
\vskip 6pt

{\bf Theorem 1:}~~{\em Let $\psi$ and $\phi$ be two vectors, and let $\rho>0$
be such that $T(\rho \hat{\x}_3)\phi\in D(i\pi/4)$ and
$T(-\rho\hat{\x}_3)\psi\in D(-i\pi/4)$.  Then for all $s\geq 0$,
\begin{eqnarray}
&&\|{\<e^{-\rho H}\psi |T(s\hat{\x}_3) Fe^{-\rho H}\phi>}|\\
&&\qquad\qquad\leq e^{-sm_0/\sqrt{2}}\!FV_3(-i\pi/4)
                   e^{-\rho H}\psi!\!FV_3(i\pi/4)e^{-\rho H}\phi!.\nonumber
\end{eqnarray}}

{\bf Proof:}  Notice that
\begin{eqnarray}
V_3(\pm i\tau)e^{-\rho H}T(\mp\rho \hat{\x}_3)V_3(\mp i\tau)
&=&T\(\pmatrix{ \cos\tau&0&0&\pm i\sin\tau\cr 0&1&0&0\cr 0&0&1&0\cr
                \pm i\sin\tau&0&0&\cos\tau\cr}
\pmatrix{i\rho\cr 0\cr 0\cr \mp\rho\cr})\\
&=&e^{-\rho(\cos\tau-\sin\tau)H}T(\mp\rho(\cos\tau+\sin\tau)\hat{\x}_3)\nonumber
\end{eqnarray}
gives a bounded operator for $0\leq\tau\leq \pi/4$.  Thus
\begin{eqnarray}
V_3(-i\tau)e^{-\rho H}\psi
&=&\({V_3(-i\tau)e^{-\rho H}T(\rho \hat{\x}_3)V_3(i\tau)})
\({V_3(-i\tau)T(-\rho \hat{\x}_3)\psi})\\
\hbox{and}&&\nonumber\\
V_3(i\tau)e^{-\rho H}\phi
&=&\({V_3(i\tau)e^{-\rho H}T(-\rho \hat{\x}_3)V_3(-i\tau)})
\({V_3(i\tau)T(\rho \hat{\x}_3)\phi})
\end{eqnarray}
can be defined for $0\leq\tau\leq \pi/4$.  If $t=\sigma+i\tau$, then
\begin{eqnarray}
T(M(t)s\hat{\x}_3)&=&T\(\pmatrix{\cosh t&0&0&\sinh t\cr 0&1&0&0\cr 0&0&1&0\cr
                                 \sinh t&0&0&\cosh t\cr}
\pmatrix{0\cr 0\cr 0\cr s\cr})\\
&=&e^{-s\cosh\sigma\sin\tau(H+P_3\tanh\sigma)}
   T\({s\cosh\sigma\cos\tau(\hat{\x}_0\tanh\sigma+\hat{\x}_3)})\nonumber
\end{eqnarray}
is also defined and bounded for $0\leq \tau\leq \pi/4$.  It follows that for
any fixed $s\geq 0$ the function
\begin{equation}
\xi(t;s)=\<V_3(t^*)e^{-\rho H}\psi|T(M(t)s\hat{\x}_3)FV_3(t)e^{-\rho H}\phi>
\end{equation}
is defined and continuous as a function of $t$ on the strip
$0\leq \tau\leq \pi/4$, and analytic on its interior.  But
$\xi(\sigma;s)=\xi(0;s)=\<e^{-\rho H}\psi|T(s\hat{\x}_3)Fe^{-\rho H}\phi>$
is constant on the real axis, and hence constant throughout the strip.  Thus
\begin{equation}
\xi(0;s)=\xi(i\tau;s)=\<V_3(-i\tau)e^{-\rho H}\psi|e^{-sH\sin\tau}
                        T(s\cos\tau\,\hat{\x}_3)FV_3(i\tau)e^{-\rho H}\phi>
\end{equation}
for any $\tau$ with $0\leq\tau\leq\pi/4$.  But since $H\geq m_0I$ on the
orthogonal complement of the vacuum, $e^{-sH\sin\tau}\leq e^{-sm_0\sin\tau}I$
on the range of $F$.  Thus
\begin{eqnarray}
\|\xi(0;s)|&\leq& e^{-sm_0\sin\tau}\!FV_3(-i\tau)e^{-\rho H}\psi!
                  \!T(s\cos\tau\,\hat{\x}_3)FV_3(i\tau)e^{-\rho H}\phi!\\
&=&e^{-sm_0\sin\tau}\!FV_3(-i\tau)e^{-\rho H}\psi!
                    \!FV_3(i\tau)e^{-\rho H}\phi!,\nonumber
\end{eqnarray}
of which (6) is the extreme case $\tau=\pi/4$.
\vskip 6pt

For example, we might choose $\psi,\phi\in R(K)+iR(K)$ for some double-cone
$K$, and let $\rho$ be such that $T(\rho\hat{\x}_3)K\subset W_R$ and
$T(-\rho\hat{\x}_3)K\subset W_L$.  Then
$T(\rho\hat{\x}_3)\phi\in D(i\pi)\supset D(i\pi/4)$, and similarly
$T(-\rho\hat{\x}_3)\psi\in D(-i\pi)\supset D(-i\pi/4)$. If we compare the
result (6) with the estimate (3), we see that in both cases there is an
exponential decay; however, the coefficient in the exponent in (6) is weaker by
a factor of $1/\sqrt{2}$.  In addition, the constant prefactor is somewhat
different.  However, in Theorem 1, the two vectors are no longer required to be
spacelike separated; their regions of localization may overlap by a distance
$\rho$, so that the two vectors $\psi$ and $\phi$ might be localized in the
same region.  If this is the case, then it is possible to interchange $\psi$
and $\phi$, and to obtain an estimate of the form
\begin{equation}
\|{\<e^{-\rho H}\psi |T(s\hat{\x}_3) Fe^{-\rho H}\phi>}|
  \leq Ce^{-\|s|m_0/\sqrt{2}}
\end{equation}
for all $s$.  We may then go further, and allow the direction to vary, and
obtain estimates of the form
\begin{equation}
\|{\<e^{-\rho H}\psi |T({\x}) Fe^{-\rho H}\phi>}|\leq Ce^{-\|x|m_0/\sqrt{2}}
\end{equation}
for a general spatial translation $T({\x})$, for some suitable $\rho$, and some
constant $C$.  It is estimates of this sort that will make possible the theorem
of the next section, which could not be established with purely
spacelike-separated estimates like (3).

\section{Free-Field Wavefunctions}

We will now specialize to the case of a free-field theory, in particular a
theory of a single free particle of mass $m$ and spin $s$.  The Hilbert space
$\H$ is a symmetric or antisymmetric Fock space based on the one-particle
Hilbert space $\h$.  When there is only a single species of particle, the
two-particle states will be symmetric for bosons, or antisymmetric for
fermions; but where there are multiplets of particles, the two-particle states
may be either spatially symmetric or spatially antisymmetric, and for this
reason everything here will be framed so as to include both cases.  Therefore
we generally suppress indications of symmetrization or antisymmetrization, and
use an unsymmetrized tensor product.  For real linear manifolds $r$, tensor
products will always be taken to be real, formed by taking limits of real
linear combinations.

The one-particle Hilbert space is $\h=\L^2(\R^3)^s$, with the action of $\PP$
on it given by
\begin{equation}
(u(a,\Lambda)f)_i({\p})=e^{ia\cdot p}
\sqrt{\omega({\p}_{\Lambda^{-1}})\over\omega({\p})}
D^{(s)}_{ij}(u_w(\Lambda;{\p}))f_j({\p}_{\Lambda^{-1}}),
\end{equation}
where $\omega({\p})=\sqrt{{\p}^2+m^2}$, $\Lambda$ is a Lorentz transform,
${\p}_\Lambda$ is the spatial part of its action on $(\omega({\p}),{\p})$, and
$u_w(\Lambda;{\p})$ is the Wigner rotation corresponding to $\Lambda$ and $\p$.
Furthermore the action of the TCP operator $\Theta$ on $\h$ will be given by
$(\Theta f)_j({\p})=e^{-i\pi j}f_{-j}^*({\p})$.  For a region $\O$ of
spacetime, we will define a real linear manifold $r(\O)$ in $\h$; then the
local algebra $\B(\O)$ will be generated by the Weyl operators corresponding to
elements of $r(\O)$, and the restriction of $R(\O)$ to the $n$-particle
subspace will be just the $n$-fold real symmetric or antisymmetric tensor
product of $r(\O)$ with itself.

We will define first of all $r(\O)$ for regions that are the causal completion
of a base subset $\O_0$ of the plane $t=0$; then from these the manifolds for
other regions can be produced by Poincar\'e transformation.  We can describe
explicitly $r(W)$ for the right wedge, whose base is a half-space, according to
the Bisognano-Wichmann result:  a wavefunction $f$ is in $r(W)$ if and only if
$f=ZU\Theta V_3(i\pi)f$, where $U$ is a rotation by angle $\pi$ about the
3-axis.  The manifolds for other such wedges can be derived from this one by
rotation and spatial translation.  If $\O_0$ is, for example, a sphere in
3-space, so that $\O$ is a double cone $K$, then $r(K)$ can be defined as the
intersection of the manifolds for the wedges whose bases contain $\O_0$.

{}From this we can see that, as stated earlier, $R(K)$ is not equal to the
intersection of the $R(W)$ for all $W\supset K$.  For example, we may consider
the two-particle component $R^{(2)}(K)=r(K)\otimes r(K)$ of $R(K)$.  Since
$r(K)$ is equal to the intersection of the $r(W)$ for all $W\supset K$, it
follows that
\begin{equation}
R^{(2)}(K)=\bigcap_{W_1,W_2\supset K}r(W_1)\otimes r(W_2),
\end{equation}
which is strictly smaller than the intersection of $R^{(2)}(W)=r(W)\otimes
r(W)$ for all $W\supset K$.  Similar results hold for the $n$-particle
components where $n>2$.  These considerations will play a large role in the
proof of the theorem to follow, in which it will be necessary to employ
translations and rotations by differing amounts in each variable separately.

Suppose $\O$ is a region with some geometrical symmetries, represented by
operators $U_i$; the action of $U_i$ on a free-field theory will be the
multiplicative promotion of the restriction $u_i$ of $U_i$ to the one-particle
space.  The modular operators $J_\O$, $\Delta_\O$ will also be multiplicative
promotions of their restrictions $j_\O$, $\delta_\O$ to the one-particle space.
But $j_\O$ and $\delta_\O$ will commute with every $u_i$; it follows that on
the $n$-particle subspace, $J_\O$ and $\Delta_\O$ will commute not only with
each $U_i$, but with every operator of the form
$u_{i_1}\otimes u_{i_2}\otimes\cdots\otimes u_{i_n}$.

Note that the one-particle Hilbert space $\h$ as a whole contains $2s+1$
vectors linearly independent up to multiplication by an overall function
$a({\p})$ of momentum---that is, up to translation.  But by the Reeh-Schlieder
principle, the translations of any manifold $r(\O)$ are total in $\h$, so each
such manifold must also contain $2s+1$ vectors linearly independent up to
multiplication by a function of momentum.

In non-relativistic quantum mechanics, one obvious feature of a (strictly)
localized wavefunction is that it is real-analytic in every component of every
momentum variable:  since it is localized in position space, by a Paley-Wiener
theorem the Fourier transform, which is the momentum-space wavefunction, is
real-analytic.  In fact, the Paley-Wiener theorem gives a precise
characterization of its analyticity, depending on the region within which it is
localized; but for now, let us consider only the fact that it is analytic in
some neighborhood of the real axes.  Since the localized wavefunctions of
relativistic quantum mechanics are not strictly localized, we cannot expect the
finer analyticity properties to carry over, but we may well ask whether these
wavefunctions are still real-analytic.

For $\O$ a compact region, it is possible to show directly that the real linear
manifold $r(\O)$ consists entirely of real-analytic functions of ${\p}$.  For
the wedge regions, although the wavefunctions are not necessarily real-analytic
functions of the momentum, it is possible to characterize them precisely based
on their analyticity in certain other variables [4].  Nevertheless, it is not
at all evident that analyticity properties of this sort will hold for the
two-particle wavefunctions $h_{jk}({\p_a},{\p_b})\in r(\O)\otimes r(\O)$, or
for more general wavefunctions in $R(\O)+iR(\O)$.

However, we can then use estimates like those of Theorem 1 to derive the
following (wherein we will for the moment ignore symmetrization and
antisymmetrization):

{\bf Theorem 2:}~~
{\em If $K$ is any double cone, then for any $h\in r(K)\otimes r(K)$,
each component $h_{jk}({\p_a},{\p_b})$ is real-analytic in the $\p_a$, $\p_b$.}

{\bf Proof:}  Without loss of generality we may, for simplicity, assume that
$K$ has its base in the plane $t=0$ and its center at the origin, with radius
$\rho$; the result for arbitrary double cones follows by Poincar\'e
transformation.  Let $f=f_j({\p_a})$ and $g=g_k({\p_b})$ be one-particle
wavefunctions associated with $r(K)$, and let us write $f_{\x}$ and $g_{\y}$
for the exponentially cut off and spatially translated wavefunctions
$f_{\x}=T(0,{\x})e^{-\rho H}f$ and $g_{\y}=T(0,{\y})e^{-\rho H}g$.  The
estimate we will use is that
\begin{equation}
\|{\<f_{\x}\otimes g_{\y}|e^{-\rho H}h>}|
\leq C e^{-m\(\|{\x}|+\|{\y}|)/\sqrt{2}}
\end{equation}
where $C$ is some positive constant (dependent on the choice of $f$ and $g$).
This will follow from Theorem 1, but let us postpone the derivation for a
moment and consider the consequences.  We can rewrite this as
\begin{equation}
\|\int d^3{\p_a}d^3{\p_b}e^{i{\x\cdot\p_a}}e^{i{\y\cdot\p_b}}
e^{-2\rho(\omega_a+\omega_b)}f_j^*({\p_a})g_k^*({\p_b})
h_{jk}({\p_a},{\p_b})|\leq Ce^{-m\(\|{\x}|+\|{\y}|)/\sqrt{2}}.
\end{equation}
The left-hand side is a Fourier transform of the function
\begin{equation}
h'({\p_a},{\p_b})=e^{-2\rho(\omega_a+\omega_b)}
f_j^*({\p_a})g_k^*({\p_b})h_{jk}({\p_a},{\p_b});
\end{equation}
thus by a Paley-Wiener theorem $h'$ is real-analytic in ${\p_a}$ and ${\p_b}$.
But $f$ and $g$ were arbitrary, save for the restriction that they be local
wavefunctions; since they are local, their components are analytic, and there
are $2s+1$ of them linearly independent up to multiplication by functions of
momentum.  This implies that each $h_{jk}$ must be real-analytic.

Let us now derive the estimate.  The methods will be those of Theorem 1, but we
will also need unitary operators defined only on the two-particle subspace,
corresponding to translations and rotations by differing amounts in each
variable.  These we will denote by $T(x;y)=T(x)\otimes T(y)$ and
$R(\theta,\hat{\x};\varphi,\hat{\y})=
R(\theta,\hat{\x})\otimes R(\varphi,\hat{\y})$; we will generally suppress
the axes and simply write $R(\theta;\varphi)$.  Since the rotations are
symmetries of $K$, we have already argued that the individual rotations
$R(\theta;\varphi)$ must commute with $J_K$ and $\Delta_K$ on the two-particle
subspace, and thus will take $r(K)\otimes r(K)$ onto itself; in other words,
$R(\theta;\varphi)\({r(K)\otimes r(K)})=
R(\theta)r(K)\otimes R(\varphi)r(K)=r(K)\otimes r(K)$.
For any particular $\x$ and $\y$, we can find an $R(\theta;\varphi)$ such that
$R(\theta)\x$ and $R(\varphi)\y$ both lie in the negative $\hat{\x}_3$
direction; thus
$R(\theta;\varphi)f_{\x}\otimes g_{\y}=
R(\theta)f_{\x}\otimes R(\varphi)g_{\y}\in e^{-\rho H}R_{W_L}$.
Then as in Theorem 1,
\begin{eqnarray}
&&\|{\<f_{\x}\otimes g_{\y}|e^{-\rho H}h>}|\nonumber\\
&&\qquad=\|{\<e^{-\rho H}f\otimes g|
          T(0,-{\x};0,-{\y})e^{-\rho H}h>}|\nonumber\\
&&\qquad=\|{\<R(\theta;\varphi)e^{-\rho H}f\otimes g|
      T(\|{\x}|\hat{\x}_3;\|{\y}|\hat{\x}_3)R(\theta;\varphi)e^{-\rho H}h>}|\\
&&\qquad=\biggl|\biggl<V_3(-i\tau)R(\theta;\varphi)
              e^{-\rho H}f\otimes g\biggm|\nonumber\\
&&\qquad\qquad e^{-(\|{\x}|\omega_a+\|{\y}|\omega_b)\sin\tau}
         T(\|{\x}|\cos\tau\,\hat{\x}_3;\|{\y}|\cos\tau\,\hat{\x}_3)
         V_3(i\tau)R(\theta;\varphi)e^{-\rho H}h\biggr>\biggr|\nonumber\\
&&\qquad\leq e^{-m(\|{\x}|+\|{\y}|)/\sqrt{2}}
        \!V_3(-i\pi/4)R(\theta;\varphi)e^{-\rho H}f\otimes g!
        \!V_3(i\pi/4)R(\theta;\varphi)e^{-\rho H}h!.\nonumber
\end{eqnarray}
Then the constant $C$ will be the supremum of the product of the two norms
above as $R(\theta;\varphi)$ varies over all individual rotations (in fact, a
suitably chosen finite set of rotations would suffice), or equivalently as the
directions of $\x$ and $\y$ vary.  This establishes the estimate (18),
and the theorem.
\vskip 6pt

This theorem clearly generalizes to the case of three or any larger number of
particles, although notational difficulties would stand in the way of writing
out the proof in the general case.  Furthermore this real analyticity is not
affected by symmetrization or antisymmetrization, nor by extension to the
(open) complex linear span.  Thus we may set it out generally, that
$n$-particle wavefunctions localized in the sense of belonging to
$R(\O)+iR(\O)$ are real-analytic functions of every component of every momentum
variable.

\section{Conclusion}

We have seen that introducing an exponential cutoff in energy enables us to
establish a cluster estimate that covers all spatial translations, not just
those which leave the observables concerned strictly spacelike separated.
Estimates of this sort have a direct application to free-field theories, in
which they can be used to establish the momentum-space analyticity of localized
wavefunctions.

\newpage
\noindent {\Large\bf References}
\medskip
\begin{enumerate}
\item H.~Araki, K.~Hepp, and D.~Ruelle, Helv. Phys. Acta 35, 164 (1962).
\item J.~J.~Bisognano and E.~H.~Wichmann, J. Math. Phys. 16, 985 (1975);
                                          J. Math. Phys. 17, 303 (1976).
\item D.~Buchholz and E.~H.~Wichmann, Commun. Math. Phys. 106, 321 (1986).
\item D.~R.~Davidson, unpublished dissertation.
\item W.~Driessler, S.~J.~Summers, and E.~H.~Wichmann,
                                     Commun. Math. Phys. 105, 49 (1986).
\item K.~Fredenhagen, Commun. Math. Phys. 97, 461 (1985).
\item R.~Haag, {\it Local Quantum Physics},  Berlin:  Springer-Verlag, 1992.
\item J.~M.~Knight, J. Math. Phys. 2, 459 (1961);
      A.~L.~Licht, J. Math. Phys. 4, 1443 (1963).
\item M.~Rieffel, Commun. Math. Phys. 39, 153 (1974).
\item R.~F.~Streater and A.~S.~Wightman,
       {\it PCT, Spin and Statistics, and All That},
        New York:  W.A. Benjamin, 1964.
\item L.~J.~Thomas and E.~H.~Wichmann,  Lett. Math. Phys. 28, 48 (1993).
\end{enumerate}

\end{document}